\pdfoutput=1

\documentclass[11pt,twoside,a4paper,cmspaper,final,collab]{cms-tdr}
\def\svnVersion{177391}\def\cmsCernNo{2012-151}\def\cmsCernDate{\today}\def\cmsMessage{Submitted to Physical Review Letters}

\begin{document}\cmsNoteHeader{SMP-12-001}

\hyphenation{had-ron-i-za-tion}
\hyphenation{cal-or-i-me-ter}
\hyphenation{de-vices}
\RCS$Revision: 140629 $
\RCS$HeadURL: svn+ssh://svn.cern.ch/reps/tdr2/papers/SMP-12-001/trunk/SMP-12-001.tex $
\RCS$Id: SMP-12-001.tex 140629 2012-07-28 09:25:26Z mpioppi $

\newcommand{\wlnu}{\PW \rightarrow \ell\Pgn}
\newcommand{\wmunu}{\PW \rightarrow \mu\Pgn}
\newcommand{\wtaunu}{\PW \rightarrow \Pgt\Pgn}
\newcommand{\wenu}{\PW \rightarrow \Pe\nu}
\newcommand{\zll}{ \cPZ/\gamma^*\rightarrow \ell^+ \ell^-}
\newcommand{\zmumu}{\cPZ/\gamma^*\rightarrow \Pgmp \Pgmm }
\newcommand{\zee}{\cPZ/\gamma^*\rightarrow \Pep\Pem }
\newcommand{\ztautau}{\cPZ/\gamma^*\rightarrow \Pgt^+ \Pgt^- }

\newcommand{\myttbar}{ \ttbar}
\newcommand{\Emt}{\ensuremath{\Em_\mathrm{T}}\xspace}
\newlength\cmsFigWidth
\ifthenelse{\boolean{cms@external}}{\setlength\cmsFigWidth{\columnwidth}}{\setlength\cmsFigWidth{0.75\textwidth}}

\cmsNoteHeader{SMP-12-001} 
\title{
\texorpdfstring{Measurement of the electron charge asymmetry in inclusive \PW\ production in \Pp\Pp\  collisions at $\sqrt{s} = 7\TeV$
}{Measurement of the electron charge asymmetry in inclusive W production in pp collisions at sqrt(s) = 7 TeV}}

\date{\today}

\abstract{
A measurement of the electron charge asymmetry
in  inclusive $\Pp\Pp \rightarrow \PW +X \rightarrow \Pe\nu + X$ production
at $\sqrt{s} = 7\TeV$ is presented based on
data
recorded by the
CMS detector at the LHC and corresponding to an integrated luminosity
of 840\pbinv. The electron charge asymmetry reflects the unequal production of \PWp\ and \PWm\ bosons in $\Pp\Pp$ collisions.
The electron charge asymmetry is measured in  bins of absolute value of electron pseudorapidity
in the range of $|\eta| < 2.4$.
The asymmetry rises from about 0.1 to 0.2 as a function of the pseudorapidity and is measured with a relative precision
better than 7\%.
This measurement provides new stringent constraints for parton distribution functions.
}

\hypersetup{%
pdfauthor={CMS Collaboration},%
pdftitle={Measurement of the electron charge asymmetry in inclusive W production in pp collisions at sqrt(s) = 7 TeV},%
pdfsubject={CMS},%
pdfkeywords={CMS, physics, electro-weak}}

\maketitle

The dominant mechanism for \PW\ boson production in $\Pp\Pp$ collisions at the Large Hadron Collider (LHC) is the annihilation of a valence quark from one proton with a sea antiquark from the other.
Because the proton contains
two valence $\cPqu$ quarks and one  valence $\cPqd$ quark,
\PWp\ bosons are produced  more frequently than \PWm\ bosons.
The Compact Muon Solenoid (CMS)  Collaboration measured  the ratio
of the inclusive \PWp\ and \PWm\ production cross sections at $\sqrt{s}= 7\TeV$
at the LHC~\cite{ewk1005}
to be in good agreement with the prediction of the standard model based on various parton
distribution functions (PDFs)~\cite{MSTW:0901, CTEQ:1007, herapdf, Ball:2011gg, Alekhin:2009ni,JimenezDelgado:2008hf}.
In this Letter we present a further investigation of inclusive \PWp\ and \PWm\ production in the $\wenu$ decay channel.
We measure the electron charge asymmetry, defined as

\newcommand{\difsig}{\ensuremath{\cPqd\sigma/\cPqd\eta}}

$$
  \mathcal{A}(\eta) = \frac{\difsig(\PWp\rightarrow\Pep\nu)-\difsig(\PWm\rightarrow\Pem\bar{\nu})}
   {\difsig(\PWp\rightarrow\Pep\nu)+\difsig(\PWm\rightarrow\Pem\bar{\nu})},
$$
where $\eta$ is the electron
pseudorapidity in the
CMS lab frame ($\eta= -\ln\tan(\theta/2)$, $\theta$ is the polar angle, measured from the anticlockwise beam direction), and
$\cPqd\sigma/\cPqd\eta$ is the differential cross section for electrons from $\PW$ boson decays.
Within the CMS tracker acceptance ($|\eta| < 2.5$), the asymmetry is expected to rise with increasing $|\eta|$ because the valence $\cPqu$ quarks that produce the
\PWp\ bosons tend to carry a higher fraction $x$ of the proton momentum~\cite{BJORKEN}
than the valence $\cPqd$ quarks that produce the \PWm\ bosons.
Measurements of the asymmetry provide constraints on the
$\cPqu$, $\cPqd$, $\cPaqu$, and $\cPaqd$ PDFs in the range $10^{-3} < x < 10^{-1}$
and are complementary to measurements of deep inelastic scattering  at HERA~\cite{herapdf}.

The lepton charge asymmetry and the $\PW$ charge
asymmetry have been studied in  $\Pp\Pap$ collisions by
the CDF and D0 Collaborations at the Fermilab Tevatron collider~\cite{CDF:wasym, D0:asym}.
The ATLAS and CMS Collaborations at the LHC reported measurements of the lepton charge asymmetry in the region $|\eta| < 2.5$
 using the data collected during the 2010 LHC runs~\cite{Aad:2011yn,Chatrchyan:2011jz}.
The LHCb Collaboration recently extended the measurement of the lepton charge asymmetry at the LHC
to the $ 2.0 < \eta < 4.5$ region ~\cite{lhcbwz}.
The NNPDF Collaboration estimated the impact of the CMS and ATLAS results on the  PDFs~\cite{Ball:2011gg};
the uncertainty on the light flavour and antiflavour distributions  is reduced by 20\% at $x \sim 10^{-3}$
and by 10--25\% at larger $x$ values.

The results presented in this Letter significantly improve the  measurement of the electron charge
asymmetry in inclusive  $\Pp\Pp \rightarrow \PW +X \rightarrow \Pe\Pgn + X$
production at
$\sqrt{s} =7\TeV$ and are based on a data sample corresponding to 840\pbinv collected in Spring 2011.
This analysis uses the portion of the 2011 dataset for which the single-electron trigger threshold was relatively low.
Compared to  the analysis performed in 2010~\cite{Chatrchyan:2011jz},
the threshold on the electron transverse momentum ($\pt$) is increased
from 25 to 35\GeV to match the higher trigger threshold for single electrons.
The data sample is ${\sim}25$ times larger than in 2010 and allows a reduction of many systematic
uncertainties.

A detailed description of the CMS experiment can be found
elsewhere~\cite{refCMS}.
The central feature of the CMS apparatus is a
superconducting solenoid of 6\unit{m} internal diameter, 13\unit{m} in length,
providing an axial field of
3.8\unit{T}. Within the field volume are the silicon pixel and strip tracker,
the crystal electromagnetic calorimeter (ECAL), and the brass/scintillator
hadron calorimeter (HCAL). Muons are measured in gas-ionization detectors
embedded in the steel return yoke of the solenoid.
The most relevant subdetectors for this measurement are the ECAL
 and the tracking system. The electromagnetic calorimeter consists
of nearly 76\,000 lead tungstate crystals, which provide coverage in pseudorapidity $|\eta| < 1.48$  in the
 barrel region and $1.48 < |\eta| < 3.00$ in two endcap regions.
A preshower detector consisting of two planes of silicon sensors interleaved with a total of $3 X_0$ of lead is located in front of the ECAL endcaps.
The energy resolution  measured for electrons from $Z$ decays varies across the acceptance of the ECAL.  It is less than 2\%
in the central region of the ECAL barrel ($|\eta|<1$), 3\% in the outer barrel, and approximately 4--5\% in the endcaps.

The  $\wenu$ candidates are characterized by a
high-$\pt$
electron accompanied by missing transverse energy $\Em_{\mathrm{T}}$,  due
to the escaping neutrino.
Experimentally,  $\Emt$ is determined as
 the magnitude of the negative vector sum of the
 transverse momenta of
all particles reconstructed using a particle-flow algorithm~\cite{pfmet} and, in this
measurement, $\Emt$ is used  to
separate signal from background on a statistical basis.
The $\wenu$ candidates used in this analysis were
collected using a set of inclusive
single-electron triggers that  did not include $\Emt$
requirements.
Multijet and photon+jet production~(QCD backgrounds)
can give rise to high-$\pt$ electron candidates and mimic $\PW$ decays.
Other background sources include Drell--Yan ($\zll$)
and $\wtaunu$ production~(electroweak or EW backgrounds),
as well as top-quark pair ($\myttbar$) production.

Monte Carlo~(MC) simulation samples are
used to develop analysis techniques and estimate some of the background contributions.
The signal and the EW backgrounds
are simulated with the
\POWHEG~\cite{POWHEG} event generator
interfaced with  the CT10~\cite{CTEQ:1007} PDF model.
The $\myttbar$  background
is generated with the \MADGRAPH~\cite{MADGRAPH} event generator
interfaced with
the CTEQ6L~\cite{CTEQ6L} PDF model.
All generated events are passed
through the  CMS detector simulation using  \GEANTfour~\cite{GEANT4}
and then processed with a reconstruction sequence identical to that used for
 collision data.

The selection criteria for electron
identification
are similar to those used in the $\PW$ and
$\cPZ$ cross section measurements~\cite{ewk1005}. A brief summary is given here
for completeness.
Electrons are identified as clusters of energy deposited in the ECAL fiducial
volume matched to tracks from the silicon tracker.
The tracks are
reconstructed using a Gaussian sum filter~(GSF)
algorithm~\cite{GSF} that takes into account possible energy loss
due to bremsstrahlung in the tracker layers. Particles misidentified  as
electrons are suppressed by requiring that the shower shape of the
ECAL cluster
be  consistent with an electron candidate,
and that the $\eta$ and $\phi$ coordinates
of the track trajectory extrapolated to the ECAL match those of the ECAL cluster.
The azimuthal angle $\phi$ is measured in the plane perpendicular to the beam axis.
Furthermore, electrons from $\PW$ decay are expected to be isolated from other activity
in the event.
We therefore require that little transverse energy be observed in the
ECAL, HCAL, and silicon tracking system within a cone
$\Delta R <$ 0.3 around the electron direction, where $\Delta R=\sqrt{(\Delta\phi)^2+(\Delta\eta)^2}$ and where the calorimeter energy deposits and the track
associated with the electron candidate are excluded.

A large fraction of electrons radiate photons,
owing to
the substantial  amount of material in front of the ECAL detector.
These photons
may convert close to the original electron trajectory, leading to
a non-negligible charge misidentification probability.
Three different methods are used to
determine the electron charge. First, the electron charge
is  determined by the signed curvature of the
associated GSF track. Second, the charge is determined from the
associated trajectory reconstructed in the silicon tracker
using a Kalman filter algorithm~\cite{KF}. Third,
the electron charge is determined from the azimuthal angle between
the vector joining
the nominal interaction point and the ECAL cluster position and the vector
joining the nominal interaction point and the innermost hit of the GSF track. It is required that  all
three charge determinations from these methods agree.
This procedure significantly reduces the charge misidentification probability
to 0.01\% in the ECAL barrel and to 0.3\%
in the ECAL endcaps.
The electron pseudorapidity is measured from the GSF track at the \Pp\Pp\ interaction point;
 the uncertainty on $\eta$ is negligible for this analysis.

The $\wenu$ candidates are selected
 by requiring electrons to have
transverse momentum
$\pt > 35\GeV$, $|\eta| < 2.4$,
 and
to be associated with one of the electron
trigger candidates used to select the electron dataset.
The Drell--Yan
and $\myttbar$ backgrounds are  suppressed
by rejecting events that contain a second isolated
electron or muon with $\pt > 15\GeV$ and $|\eta| < 2.4$.
According to MC simulations,
the data sample of selected electrons
consists of
 about 16\% QCD background events,
about
 7.4\%  EW background
events,
and about 0.4\% $\myttbar$ background events.

As the forward and backward hemispheres are equivalent in \Pp\Pp\ collisions,
the asymmetry results are presented as a function of the absolute value
of the electron pseudorapidity.
The selected events are divided into bins of $|\eta|$ with bin width  0.2.
The region of $|\eta|$ between 1.4 and 1.6 is excluded
because the cables and services in the transition region
between the ECAL barrel and endcaps cause a significant
reduction in the efficiency and purity of the $\PW$ selection.
A total of 1\,229\,315 $\PWp\rightarrow \Pep\Pgn$ candidates and
991\,256 $\PWm\rightarrow \Pem\cPagn$ candidates
are selected in the 11 bins of $|\eta|$.

\begin{figure}[tbh]
  \begin{center}
    \includegraphics[width=\columnwidth]{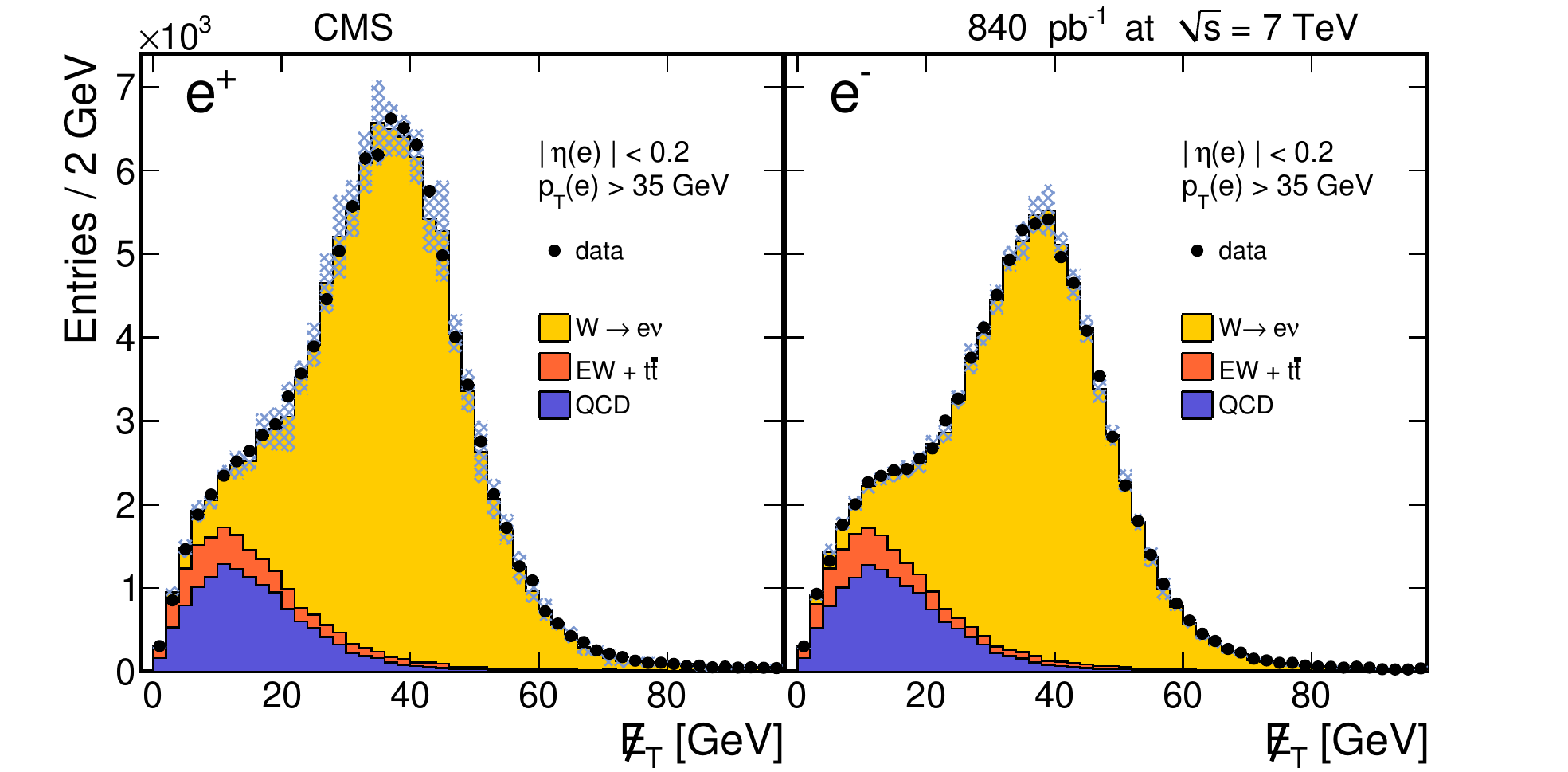}
    \includegraphics[width=\columnwidth]{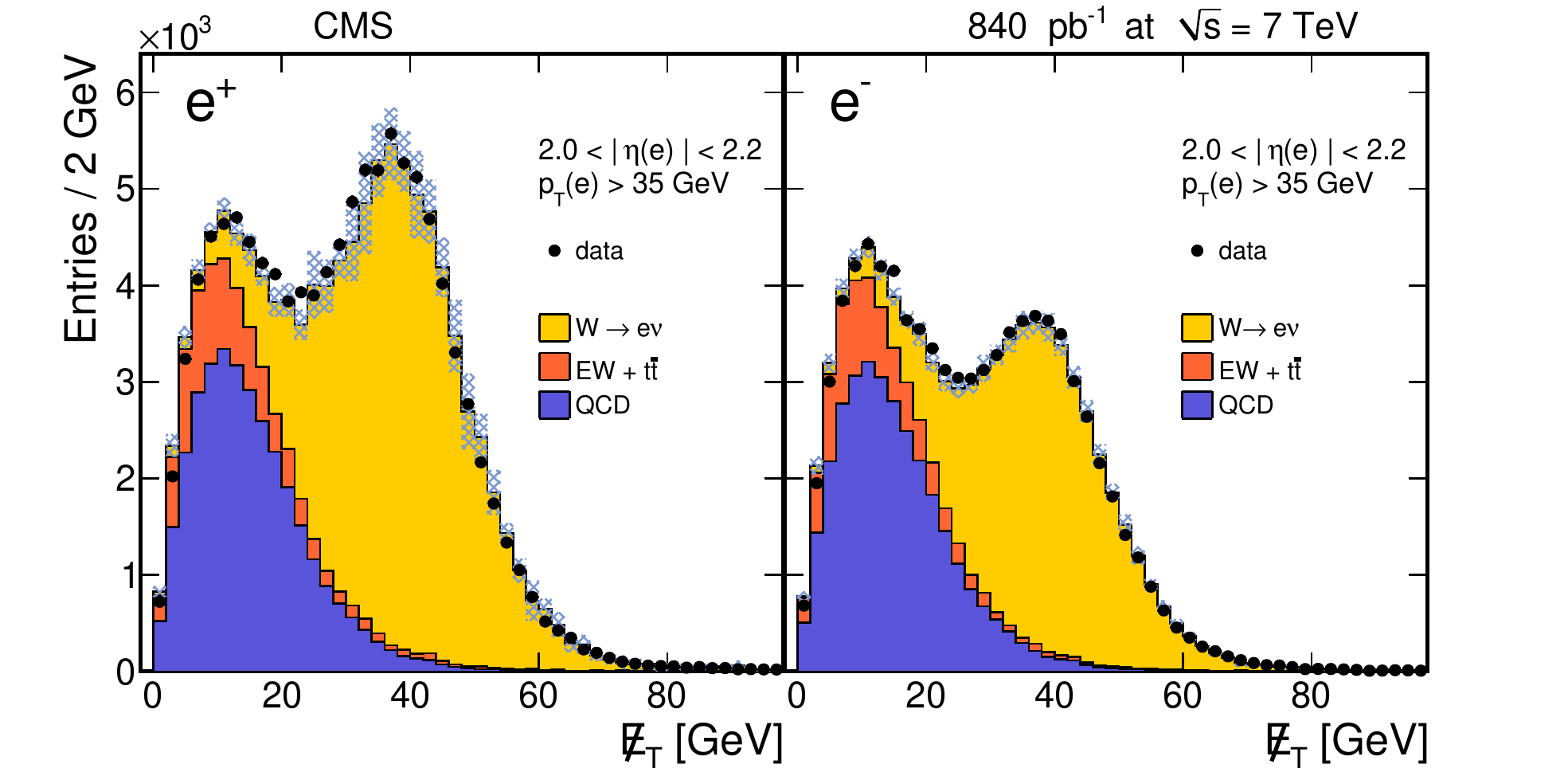}
    \caption{Signal plus background fit to data $\Emt$  distributions for positrons (left) and electrons (right) in data.
Results are shown for
the first pseudorapidity bin ($|\eta| < 0.2$, top) and for the tenth pseudorapidity bin
($2.0 < |\eta| < 2.2$, bottom).
The hatched area represents the statistical and systematic uncertainties associated with the fitting procedure.}
    \label{fig1}
  \end{center}
\end{figure}

A binned extended maximum-likelihood fit is performed
to the $\Emt$ distribution to estimate the $\wenu$ signal
yield for electrons~($N^-$) and positrons~($N^+$) in each pseudorapidity bin.
The signal $\Emt$ shape is derived
from MC simulations with an event-by-event correction to account for differences in energy
scale and resolution between data and simulation inferred from
 the hadronic recoil energy distributions in $\zee$ events selected
from data~\cite{RECOIL}.
 The shape of the QCD background is
determined, for each charge, from a signal-free control sample obtained by
inverting a subset of the electron identification criteria~\cite{ewk1005}.
The $\Emt$ shapes for
other backgrounds such as the Drell--Yan process,
 $\myttbar$,
 and $\wtaunu$ are taken from MC simulations with
a fixed  normalization relative to the $\wenu$
yields. The normalization factors are calculated from
the predicted values of the cross sections at next-to-leading order (NLO).
The yield of the  QCD  background and the
 yield of the $\wenu$ signal ($N^\pm$) are free parameters
in the fit. The results of the fits to the data
are shown for the first ($|\eta| < 0.2$) and the tenth ($2.0 < |\eta| < 2.2$) pseudorapidity bins in
Fig.~\ref{fig1}.
The uncertainty in each bin  represents the systematic and statistical
components associated with the $N^\pm$ extraction procedure.
The ratio of the QCD background yields estimated from the fit 
to the $\Em_{\mathrm{T}}$ distributions for positive and 
negative electrons is consistent with unity.
The EW background yield for positive electrons is larger 
than for negative electrons because of the contribution of $\wtaunu$ events.
The charge asymmetry is
obtained from the
$(N^+-N^-)/(N^++N^-)$ ratio.

Two sources of systematic uncertainty are related to the signal $\Emt$
shape, the PDF used to generate the events and the uncertainty on the
correction from the hadronic recoil applied to the signal $\Emt$ shape.
The PDF uncertainties and their effects on the measured asymmetries are evaluated using
the prescription given by the
PDF4LHC Working Group~\cite{Botje:2011sn}.
Uncertainties on the correction from the hadronic recoil obtained from  data are also propagated
to the asymmetry measurements.
The systematic uncertainty due to the QCD background shape is evaluated by studying samples from different QCD background control regions.
The systematic uncertainty due to the modelling of Drell--Yan,  $\myttbar$,
 and $\wtaunu$ is estimated by varying the relative normalization of the EW backgrounds
to the $\wenu$
yield by the uncertainty on the Drell--Yan and $\myttbar$ cross sections,
and the  effect on the observed asymmetry is negligible.
The values of $N^\pm$ from the fitting procedure are insensitive to the presence
of pile-up interactions
because the $\Emt$ distributions are obtained from data.

In order to compare our results directly
to theoretical predictions, the observed charge asymmetry is corrected
for three detector effects: 1) electron energy scale and resolution,
2) relative detection efficiency of positrons and electrons, and 3) electron charge
misidentification.

The electron energy scale and resolution can bias the asymmetry
because of the effect on electrons with transverse momentum
close to the threshold value of 35\GeV.
The electron energy scale and resolution are  determined directly from  the $\zee$
data and are used to adjust the simulated electron energy at the generator level. The correction to
the measured
charge asymmetry  is estimated in each pseudorapidity bin
by comparing the charge asymmetry as determined in the simulation
with the resulting
asymmetry after smearing.
The corrections for the electron energy scale and resolution are found to fall between $-4.4 \times 10^{-3}$ and $0.2 \times 10^{-3}$.
The  uncertainties on the energy scale and
resolutions are taken as sources for
systematic uncertainties.
The charge asymmetry is also corrected for final-state radiation,  and
the uncertainty  on the correction is taken as an additional
systematic uncertainty, which is summed in quadrature with
 the uncertainty on the electron energy scale and resolution.

Any efficiency difference between electrons and positrons
would bias
 the measured charge asymmetry.
The total electron efficiency (including electron reconstruction, identification,
and trigger efficiencies) in each pseudorapidity bin is measured
using the $\zee$ data, separately for \Pem\ and
\Pep, using the tag-and-probe method~\cite{ewk1005}.
The efficiency ratio is calculated and
found to be between 0.96 and 1.03;
the statistical uncertainties on the efficiency ratios (0.01--0.03) are treated
as systematic uncertainties.
This is
the dominant systematic uncertainty in all the pseudorapidity bins.

The true charge asymmetry $\mathcal{A}$  is diluted
 because of charge misidentification, resulting in an observed asymmetry $\mathcal{A}^{\text{obs}}= \mathcal{A}(1-2w).$
The electron charge misidentification probability $w$ is measured
using $\zee$ events in data.
The observed electron charge asymmetry is corrected for the charge
misidentification probability as a function of $|\eta|$.
The statistical uncertainty  on the
charge misidentification probability is taken as the systematic
uncertainty.

Table~\ref{tab1} summarizes the systematic uncertainties in all the electron pseudorapidity bins.
The full systematic covariance matrix is given in Table~\ref{tabcorr}.

\begin{table}[htbp]
    \topcaption{\label{tab1}Summary of the systematic uncertainties on the asymmetry. All values are given in units of $10^{-3}$. }
    \begin{center}
      \begin{tabular}{c|cccc}
      $|\eta|$ bin   &Signal & Energy & Charge &  Efficiency \\
        & Yield & Scale \& Res. & MisId. & Ratio \\ \hline
  $0.0<|\eta|<0.2$ & 1.8 & 0.6 & $<$0.1 &  4.5 \\
  $0.2<|\eta|<0.4$ & 2.5 & 0.6 & $<$0.1 &  4.4 \\
  $0.4<|\eta|<0.6$ & 2.7 & 0.3 & $<$0.1 &  4.4 \\
  $0.6<|\eta|<0.8$ & 2.5 & 0.3 & $<$0.1 &  4.4 \\
  $0.8<|\eta|<1.0$ & 1.9 & 0.6 & 0.1 &  4.4 \\
  $1.0<|\eta|<1.2$ & 2.4 & 1.0 & 0.1 &  4.9 \\
  $1.2<|\eta|<1.4$ & 2.6 & 0.8 & 0.1 &  5.4 \\
  $1.6<|\eta|<1.8$ & 3.1 & 0.8 & 0.1 &  9.2 \\
  $1.8<|\eta|<2.0$ & 2.0 & 1.6 & 0.2 &  8.7 \\
  $2.0<|\eta|<2.2$ & 2.0 & 2.6 & 0.3 & 10.0 \\
  $2.2<|\eta|<2.4$ & 2.9 & 2.4 & 0.3 & 12.5 \\
      \end{tabular}
    \end{center}
  \end{table}

\begin{table*}[htbp]
 \topcaption{\label{tabcorr}Covariance matrix for  the systematic uncertainties on the asymmetry.
                               All values are given in units of $10^{-6}$.
The matrix is symmetric.}
   \begin{center}
      \begin{tabular}{l|ccccccccccc}
$|\eta|$ bin& \tiny{$\left[0.0,0.2\right]$}
& \tiny{$\left[0.2,0.4\right]$}
& \tiny{$\left[0.4,0.6\right]$}
& \tiny{$\left[0.6,0.8\right]$}
& \tiny{$\left[0.8,1.0\right]$}
& \tiny{$\left[1.0,1.2\right]$}
& \tiny{$\left[1.2,1.4\right]$}
& \tiny{$\left[1.6,1.8\right]$}
& \tiny{$\left[1.8,2.0\right]$}
& \tiny{$\left[2.0,2.2\right]$}
& \tiny{$\left[2.2,2.4\right]$} \\ \hline

$\left[0.0,0.2\right]$ & 23.7 & &  &  & &  &  & & &  &  \\
$\left[0.2,0.4\right]$ & 2.6 & 26.2 & & &  &  &  &  &  &  &  \\
$\left[0.4,0.6\right]$ & 2.2 & 2.6 & 26.6 &  &  &  &  &  &  & &  \\
$\left[0.6,0.8\right]$ & 2.5 & 2.9 & 2.6 & 25.6 &  &  &  &  &  & &  \\
$\left[0.8,1.0\right]$ & 2.7 & 3.1 & 2.8 & 3.3 & 23.3 &  &  &  &  &  &  \\
$\left[1.0,1.2\right]$ & 2.9 & 3.3 & 2.9 & 3.4 & 3.9 & 30.8 &  &  &  &  &  \\
$\left[1.2,1.4\right]$ & 2.9 & 3.4 & 3.2 & 3.7 & 4.2 & 4.5 & 36.5 &  &  &  &  \\
$\left[1.6,1.8\right]$ & 2.9 & 3.2 & 2.9 & 3.3 & 3.7 & 4.1 & 4.3 & 94.9 &  &  &  \\
$\left[1.8,2.0\right]$ & 2.8 & 3.0 & 2.5 & 2.8 & 3.4 & 4.0 & 3.7 & 3.8 & 82.4 &  &  \\
$\left[2.0,2.2\right]$ & 3.1 & 3.9 & 3.3 & 3.7 & 4.6 & 5.7 & 5.8 & 5.1 & 6.2 & 110.7 &  \\
$\left[2.2,2.4\right]$ & 4.2 & 4.7 & 4.1 & 4.7 & 5.6 & 6.8 & 6.7 & 6.2 & 7.0 & 10.3 & 171.0 \\
    \end{tabular}
    \end{center}
\end{table*}
The measured charge asymmetry results are summarized in
Table~\ref{tab2}
with both statistical
and systematic uncertainties shown.
The statistical uncertainties in the various pseudorapidity bins are uncorrelated.

\begin{figure}[htbp]
  \begin{center}
    \includegraphics[width=\cmsFigWidth]{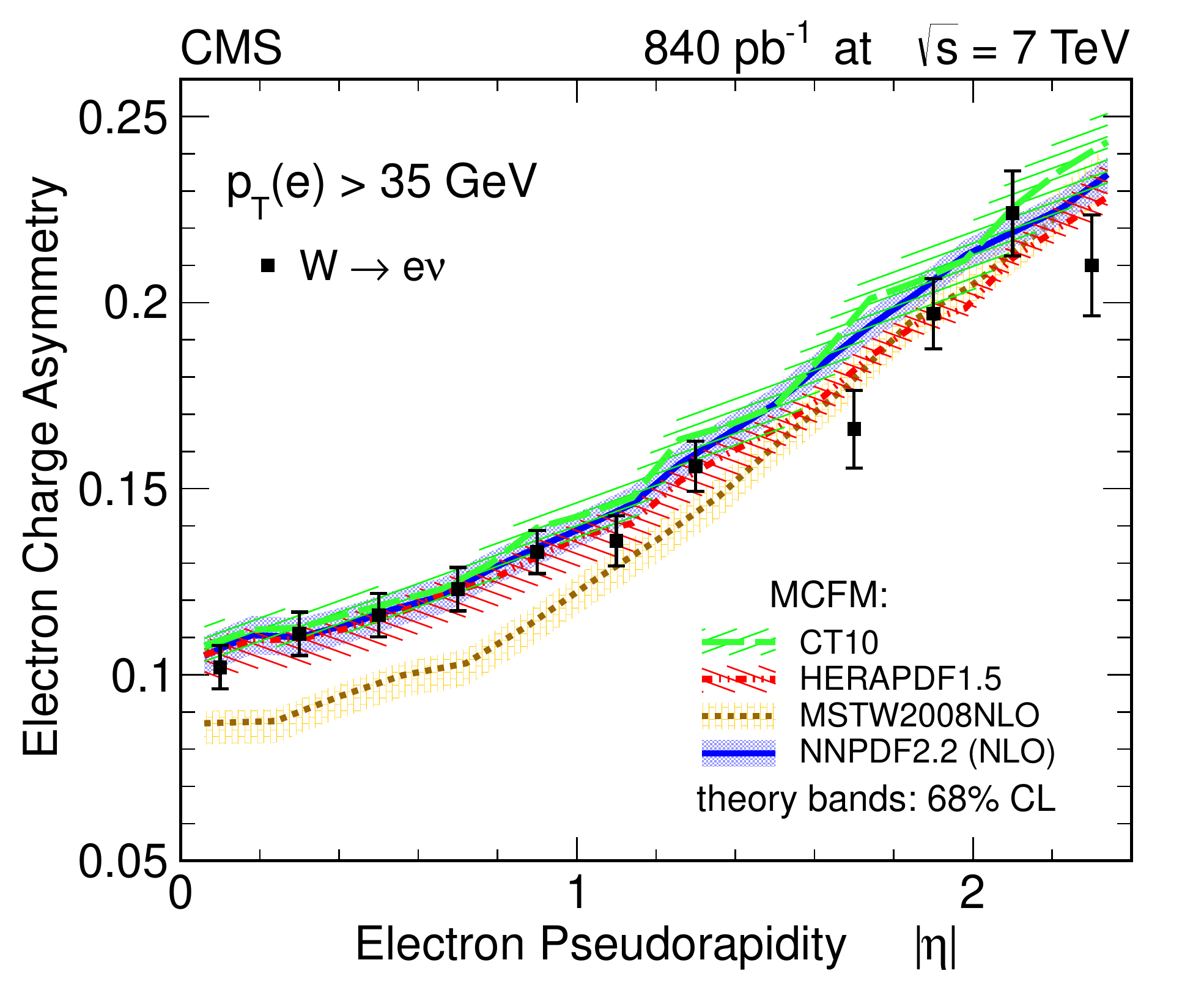}
    \caption{Comparison of the measured electron
asymmetry to the predictions of
different PDF models for electron $\pt >$ 35\GeV.
The error bars include both statistical and systematic uncertainties.
The data points are placed in the center of the $|\eta|$ bins.
The PDF uncertainty bands are estimated using the PDF reweighting technique and correspond to 68\% confidence level.
}
    \label{fig2}
  \end{center}	
\end{figure}

The experimental results are compared in Table~\ref{tab2} and in Fig.~\ref{fig2} to theoretical predictions
obtained with the NLO \MCFM~\cite{MCFM} generator interfaced with
CT10~\cite{CTEQ:1007}, HERAPDF~\cite{herapdf},
NNPDF~\cite{Ball:2011gg}, and  MSTW2008NLO~\cite{MSTW:0901} PDF models.
The theoretical errors  are estimated using
the PDF reweighting technique~\cite{PDFERROR}.
The experimental data are in agreement with the predictions from CT10, NNPDF, and HERAPDF, while the predictions from MSTW are
systematically lower than the observed asymmetry in the region $|\eta|<1.4$.

\begin{table}[htbp]
    \topcaption{\label{tab2}Summary of the measured charge asymmetry results.
        The first uncertainty is
statistical and the second is systematic. The theoretical predictions are
obtained using \MCFM interfaced with four different PDF models.
 The PDF uncertainties are estimated using the PDF reweighting
technique.
All values are
 in units of $10^{-3}$.
}
    \begin{center}
      \begin{tabular}{l|l|cccc}
$|\eta|$ bin& Measured & \multicolumn{4}{c} {Theoretical Predictions} \\
 & Asymmetry $\mathcal{A}$ & CT10 & HERAPDF & MSTW &NNPDF \\ \hline
  $\left[0.0,0.2\right]$ &$102\pm3\pm5$ &$109\pm5$ &$106^{+4}_{-8}$ & $87^{+3}_{-5}$&$ 107\pm5$\\
  $\left[0.2,0.4\right]$ &$111\pm3\pm5$ &$114\pm5$ &$110^{+4}_{-8}$ & $89^{+3}_{-5}$&$ 110\pm5$\\
  $\left[0.4,0.6\right]$ &$116\pm3\pm5$ &$119\pm5$ &$115^{+4}_{-8}$ & $98^{+3}_{-5}$&$ 116\pm5$\\
  $\left[0.6,0.8\right]$ &$123\pm3\pm5$ &$126\pm5$ &$122^{+4}_{-8}$ & $103^{+3}_{-5}$&$ 123\pm5$\\
  $\left[0.8,1.0\right]$ &$133\pm3\pm5$ &$138^{+5}_{-6}$ &$132^{+4}_{-8}$ & $115^{+4}_{-5}$&$ 134\pm5$\\
  $\left[1.0,1.2\right]$ &$136\pm3\pm6$ &$146\pm6$ &$140^{+5}_{-8}$ & $128^{+4}_{-5}$&$145\pm5$ \\
  $\left[1.2,1.4\right]$ &$156\pm3\pm6$ &$164^{+6}_{-7}$ &$153^{+5}_{-7}$ & $144\pm5$&$158\pm5$ \\
  $\left[1.6,1.8\right]$ &$166\pm3\pm10$ &$195^{+8}_{-9}$ &$181\pm5$ & $179\pm5$&$190\pm4$ \\
  $\left[1.8,2.0\right]$ &$197\pm3\pm9$ &$207^{+8}_{-10}$ &$196^{+4}_{-3}$ & $200^{+6}_{-5}$&$206\pm4$ \\
  $\left[2.0,2.2\right]$ &$224\pm3\pm11$ &$224^{+8}_{-11}$ &$211^{+5}_{-3}$ & $213^{+6}_{-5}$&$219\pm4 $\\
  $\left[2.2,2.4\right]$ &$210\pm4\pm13$ &$241^{+8}_{-12}$ &$225^{+9}_{-4}$ & $231^{+6}_{-5}$&$231\pm5 $\\
      \end{tabular}
    \end{center}
  \end{table}

In summary, we have measured the electron charge asymmetry in the
$\wenu$ channel in a sample of proton-proton collisions at 7\TeV,
corresponding to an integrated luminosity of 840\pbinv.
The measured asymmetry rises from about 0.1 to 0.2 as a function of the pseudorapidity,
with an uncertainty that ranges from 0.006 in the central region
to 0.014 in the ECAL endcaps.
This precise measurement of the electron charge asymmetry
in inclusive $\PW$ production at the LHC
provides stringent constraints for parton distribution functions.

We congratulate our colleagues in the CERN accelerator departments for the excellent performance of the LHC machine. We thank the technical and administrative staff at CERN and other CMS institutes, and acknowledge support from: FMSR (Austria); FNRS and FWO (Belgium); CNPq, CAPES, FAPERJ, and FAPESP (Brazil); MES (Bulgaria); CERN; CAS, MoST, and NSFC (China); COLCIENCIAS (Colombia); MSES (Croatia); RPF (Cyprus); MoER, SF0690030s09 and ERDF (Estonia); Academy of Finland, MEC, and HIP (Finland); CEA and CNRS/IN2P3 (France); BMBF, DFG, and HGF (Germany); GSRT (Greece); OTKA and NKTH (Hungary); DAE and DST (India); IPM (Iran); SFI (Ireland); INFN (Italy); NRF and WCU (Korea); LAS (Lithuania); CINVESTAV, CONACYT, SEP, and UASLP-FAI (Mexico); MSI (New Zealand); PAEC (Pakistan); MSHE and NSC (Poland); FCT (Portugal); JINR (Armenia, Belarus, Georgia, Ukraine, Uzbekistan); MON, RosAtom, RAS and RFBR (Russia); MSTD (Serbia); MICINN and CPAN (Spain); Swiss Funding Agencies (Switzerland); NSC (Taipei); TUBITAK and TAEK (Turkey); STFC (United Kingdom); DOE and NSF (USA).

\bibliography{auto_generated}
\cleardoublepage \appendix\section{The CMS Collaboration \label{app:collab}}\begin{sloppypar}\hyphenpenalty=5000\widowpenalty=500\clubpenalty=5000\textbf{Yerevan Physics Institute,  Yerevan,  Armenia}\\*[0pt]
S.~Chatrchyan, V.~Khachatryan, A.M.~Sirunyan, A.~Tumasyan
\vskip\cmsinstskip
\textbf{Institut f\"{u}r Hochenergiephysik der OeAW,  Wien,  Austria}\\*[0pt]
W.~Adam, T.~Bergauer, M.~Dragicevic, J.~Er\"{o}, C.~Fabjan\cmsAuthorMark{1}, M.~Friedl, R.~Fr\"{u}hwirth\cmsAuthorMark{1}, V.M.~Ghete, J.~Hammer, N.~H\"{o}rmann, J.~Hrubec, M.~Jeitler\cmsAuthorMark{1}, W.~Kiesenhofer, V.~Kn\"{u}nz, M.~Krammer\cmsAuthorMark{1}, D.~Liko, I.~Mikulec, M.~Pernicka$^{\textrm{\dag}}$, B.~Rahbaran, C.~Rohringer, H.~Rohringer, R.~Sch\"{o}fbeck, J.~Strauss, A.~Taurok, P.~Wagner, W.~Waltenberger, G.~Walzel, E.~Widl, C.-E.~Wulz\cmsAuthorMark{1}
% ...
\vskip\cmsinstskip
\dag:~Deceased\\
1:~~Also at Vienna University of Technology, Vienna, Austria\\
% ...

\end{sloppypar}
\end{document}